\documentclass[final,conference,twocolumn,letterpaper]{IEEEtran}

\IEEEoverridecommandlockouts              
\usepackage{fancyhdr}
\usepackage{graphicx}
\usepackage[utf8]{inputenc}

\graphicspath{{figures/}}
\usepackage{url}
\usepackage{wrapfig}
\usepackage{units}
\usepackage{caption}
\usepackage{acro}
\RequirePackage{xstring}
\RequirePackage{xparse}
\NewDocumentCommand\acrodef{mO{#1}mG{}}{\DeclareAcronym{#1}{short={#2}, long={#3}, #4}}
\acrodef{ap}[AP]{access point}
\acrodef{RF}{radio frequency}
\acrodef{VLC}{Visible Light Communication}
\acrodef{LoS}{line-of-sight}
\acrodef{NLoS}{non-line-of-sight}
\acrodef{MIMO}{Multiple-Input Multiple-Output}
\acrodef{COTS}{commercially available off-the-shelf}
\acrodef{CSI}{channel state information}
\acrodef{IoT}{internet of things}
\acrodef{LPWAN}{low-power wide area network}

\usepackage{calc}
\usepackage{graphicx}
\usepackage[lined,boxed, vlined,ruled,linesnumbered]{algorithm2e}
\usepackage{amsmath,amssymb,amsthm,amsfonts,amsbsy}
\usepackage{gensymb}

\usepackage{algpseudocode}
\usepackage{color}
\usepackage[table]{xcolor} % DVI only
\usepackage[caption=false]{subfig}
\usepackage{dblfloatfix}
\usepackage{nicefrac}
\usepackage{tcolorbox}
\usepackage{txfonts}

\usepackage{multirow}

\newcolumntype{C}[1]{>{\centering\arraybackslash}p{#1}}

\usepackage{booktabs}

\definecolor{BgGray}{gray}{0.7}%
\definecolor{BgGray2}{gray}{0.96}%
\definecolor{RowColorOdd}{named}{BgGray2}%
\definecolor{RowColorEven}{named}{white}%
\definecolor{comments}{gray}{.5}
\definecolor{Gray}{gray}{0.85}

\DeclareMathOperator*{\argmax}{argmax} % thin space, limits underneath in displays
\providecommand{\keywords}[1]{\textbf{\textit{Index terms---}} #1}

\usepackage{pifont}% http://ctan.org/pkg/pifont
\usepackage[multiuser,nomargin,marginclue,footnote]{fixme}
\fxusetheme{color}
\fxsetup{targetlayout=colorcb}
\FXRegisterAuthor{all}{anall}{ALL}
\FXRegisterAuthor{md}{anmd}{MD}
\FXRegisterAuthor{tolja}{antolja}{TOLJA}
\FXRegisterAuthor{pg}{anpg}{PG}
\FXRegisterAuthor{mch}{anmc}{MC}
\FXRegisterAuthor{cp}{ancp}{CP}
\FXRegisterAuthor{awo}{anawo}{AWO}
\fxsetup{status=draft}

\begin{document}
\providetoggle{techreport}
\settoggle{techreport}{false}

% new \todo{...} command
\def\todo{%
	% start drawing bar in margin
	\cbcolor{magenta}
	\cbstart%
	% open curly brace
	\begingroup
	\color{magenta}
	% line breaks are line breaks...
	\obeylines%
	% ...paragraphs are paragraphs...
	\begingroup\lccode`~=`\^^M\lowercase{\endgroup\def~}{\par\leavevmode}%
	% ...underscore is underscore...
	\catcode`\_=\active
	% ...and so are <>#^
	\catcode`\<=\active\lccode`~=`<\lowercase{\def~}{$<$}%
	\catcode`\>=\active\lccode`~=`>\lowercase{\def~}{$>$}%
	\catcode`\#=\active\lccode`~=`\#\lowercase{\def~}{$\#$}%
	\catcode`\^=\active\lccode`~=`\^\lowercase{\def~}{$\hat{~}$}%
	% don't toke too early
	\todoCtd
}\def\todoCtd#1{%
	TODO: #1%
	% if nothing was to be printed, print three dots
	\ifx&#1&...\fi%
	% now forget everything
	\endgroup
	% stop drawing bar in margin
	\cbend
	% stop accepting tokens for this command
	\relax
}

\newcommand{\proposalName}{\texttt{Wi-Lo}}

\title{\proposalName{}: Emulating LoRa using COTS WiFi}

\author{
\IEEEauthorblockN{Piotr Gaw{\l}owicz, Anatolij Zubow, Falko Dressler}
\IEEEauthorblockA{Technische Universität Berlin, Germany}
\{gawlowicz, zubow, dressler\}@tkn.tu-berlin.de
}

\maketitle

\begin{abstract}
We present Wi-Lo, which allows to convert an ordinary 802.11 (WiFi) access point into an \ac{IoT} gateway supporting the \ac{LPWAN} technology LoRa in the downlink.
Our Wi-Lo system only requires a software update and no additional hardware.
It uses signal emulation technique based on complementary code keying modulation from 802.11b in order to emulate a downlink LoRa (long range) transmission.
The Wi-Lo gateway can be used by a normal WiFi-enabled smartphone to send packets to LoRa compliant IoT devices like smart sensors.
We implemented a prototype using commodity WiFi hardware.
Experimental results show that Wi-Lo enables a normal WiFi node to communication to LoRa devices even over long distances, which is comparable to the configurations using pure LoRa transmitter and receivers.
\end{abstract}


% -------------- Section end marker --------------
%                _       _
%               ( )_    ( )
%    ___  _   _ | ,_)   | |__     __   _ __   __
%  /'___)( ) ( )| |     |  _ `\ /'__`\( '__)/'__`\
% ( (___ | (_) || |_    | | | |(  ___/| |  (  ___/
% `\____)`\___/'`\__)   (_) (_)`\____)(_)  `\____)
%
% -------------- Section end marker --------------

\acresetall
\section{Introduction}\label{sec:intro}
Today, we see a constant growth in the number of connected devices forming the \ac{IoT} idea.
\Acp{LPWAN} are an attractive way to connect such a large number of IoT devices.
LPWANs enable low-power, often battery-powered, devices to communicate wirelessly over long distances but at very low data rates.
Among many LPWANs, Long Range Wide Area Network (LoRaWAN)~\cite{sornin2015lorawan} becomes a widely used technology, which also has attracted many interests from research and academia.
Usually LoRa radios are operating on sub-gigahertz spectrum, however, recently a new so-called 2.4\,GHz LoRa becomes available which uses the globally harmonized 2.4\,GHz ISM band~\cite{rander-andersen2020ranging}.
The key benefits are the larger available spectrum in 2.4\,GHz, i.e. 80\,MHz compared to just a few MHz in sub-GHz, which allows to operate multiple LoRa channels in parallel as well as the not required strict channel duty cycling.
Moreover, the maximum available bandwidth is increased to 1.6\,MHz, resulting in higher data rate.
This allows 2.4\,GHz LoRa to support a wider range of IoT applications. % requiring higher data rates.

In order to connect to a 2.4\,GHz LoRa device using standard smartphone or tablet a multi-technology gateway (MTG) is required for translating WiFi or Bluetooth packets to 2.4\,GHz LoRa packets.
However, such a MTGs increase the cost of operation.
In this paper, we present Wi-Lo, which uses signal emulation technique to emulate a 2.4\,GHz LoRa transmission using commodity 802.11 (WiFi) hardware.
Therefore, Wi-Lo exploits the single carrier complementary code keying (CCK) waveform used by the IEEE 802.11b standard for the emulation.
This is achieved by carefully selecting the WiFi payload and hence the used CCK chipping sequence to emulate the LoRa waveform as close as possible.
Hence, with Wi-Lo it becomes possible to convert a commodity residential or enterprise WiFi AP into a multi-technology gateway with a simple software update and no additional hardware.
Such a Wi-Lo enabled WiFi AP can communicate in the downlink using both normal WiFi as well as 2.4\,GHz LoRa transmissions using a single WiFi COTS network interface card.
Our work is complementary to XFi approach~\cite{liu2020xfi}, which enables the reception of LoRa using WiFi whereas Wi-Lo targets LoRa transmissions using COTS WiFi hardware. 
Wi-Lo was prototypically implemented, tested and evaluated.
Experimental results show that Wi-Lo enables a normal WiFi node to communication to LoRa devices even over a long distance, which is comparable to configurations using pure LoRa transmitter and receivers.

% -------------- Section end marker --------------
%                _       _
%               ( )_    ( )
%    ___  _   _ | ,_)   | |__     __   _ __   __
%  /'___)( ) ( )| |     |  _ `\ /'__`\( '__)/'__`\
% ( (___ | (_) || |_    | | | |(  ___/| |  (  ___/
% `\____)`\___/'`\__)   (_) (_)`\____)(_)  `\____)
%
% -------------- Section end marker --------------

\section{Background}\label{sec:background}
This section gives a detailed overview of the physical layers of the 2.4\,GHz LoRa and IEEE 802.11b technology.

\begin{figure}
\centering
\includegraphics[width=0.80\linewidth]{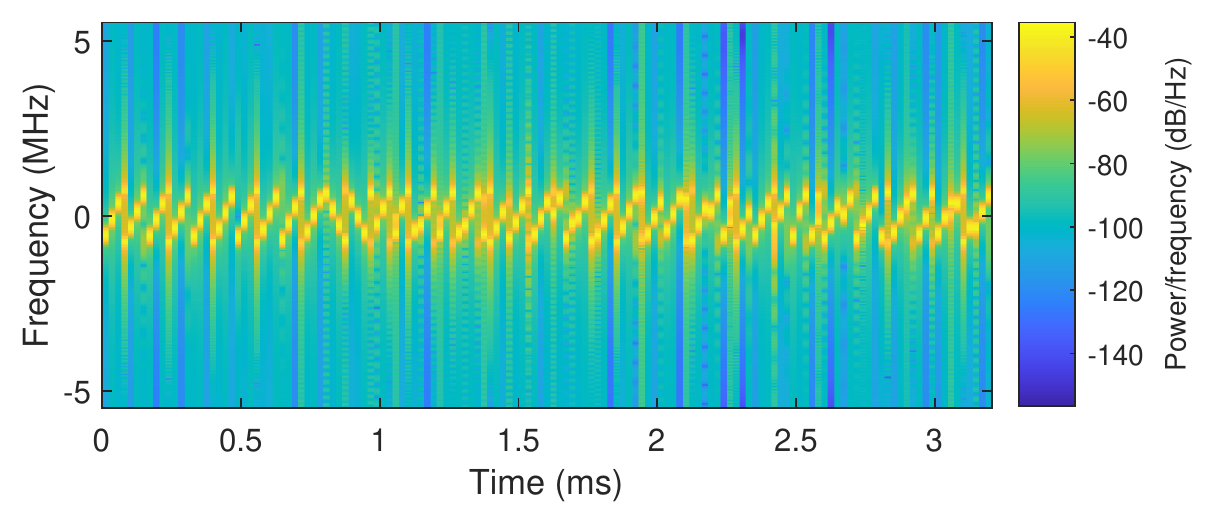}
%\vspace{-5pt}
\caption{Spectrum of LoRa transmission (1.6\,MHz, SF7).}
%\vspace{-0.5em}
\label{fig:lora_spectrum1}
\end{figure}

\subsection{LoRa PHY in 2.4\,GHz}
Long Range (LoRa) technology was developed by Semtech~\cite{seller2016new}.
The modulation scheme used in 2.4\,GHz LoRa is Chirp Spectrum Spread (CSS), which is the same as for sub-GHz LoRa (433/868\,MHz for EU).
%
%A chirp is a sinusoidal signal of frequency increases or decreases over time.
%
CSS modulation produces a chirp signal where all chirps have practically the same time duration (Fig.~\ref{fig:lora_spectrum1}). 
A chirp is characterized by a time profile of the instantaneous frequency that changes over the time interval $T$ from a frequency $f_0$ to $f_1$~\cite{haxhibeqiri2018survey}.
Moreover, it uses the entire bandwidth, making it robust to noise and interference and therefore can be received at even very low power.
There are two types of different chirps~\cite{haxhibeqiri2018survey}: the base chirp whose frequency time profile starts with the minimal frequency $f_{\min} = -\mathrm{BW}/2$ and ends with the maximal frequency $f_{\max} = +\mathrm{BW}/2$, where BW being the spreading bandwidth of the signal. %, and modulated chirps that are cyclically time shifted base chirps.
The chirp that start with frequency $f_{\max}$ and ends with $f_{\min}$ is referred to as a down-chirp. 
2.4\,GHz LoRa provides a wider bandwidth than sub-GHz LoRa, i.e., up to 1600\,kHz as compared to just 500\,kHz, resulting in a larger data rate. %\todo{if the chirps sweep through the spectrum, how will more BW enable more data rate}
%
%This is higher compared to sub-GHz LoRa which provides a bandwidth of only up to 500\,kHz. % resulting in data rate of 21.875\,kbps.
%
%The time shift of each chirp at the receiver side can be determined after the alignment of the time reference between receiver and transmitter by means of preamble detection.
%
To further, improve the robustness against noise and interference, LoRa uses diagonal interleaving as well as forward error correction (FEC) codes with code rates from 4/5 to 4/8. 
%
%The data rate and symbol rate depends on the SF and the bandwidth used.
%
%The symbol rate is given by the formula
%
Moreover, dynamic Spreading Factor (SF) is used to trade data rate for sensitivity.
The packet structure at the physical layer includes a preamble, an optional header and the data payload~\cite{haxhibeqiri2018survey}.
The preamble, which starts with a sequence of constant upchirp symbols (cf. Fig.~\ref{fig:lora_spectrum1}), is used to synchronize the receiver with the transmitter. % and can have a variable length.
%
%The fixed part of the preamble consists of four symbols, and the rest is programmable with a minimal length of six symbols and a maximal length of 65,532.
%
%The preamble starts with a sequence of constant upchirp symbols that is programmable and helps to detect the start of the frame.
%
The preamble is followed by two chirp symbols encoding the sync word, which is used for frame synchronization.
%
%Usually, the two chirps will modulate opposite values.
%
%The sync word can also be used to distinguish between devices from different networks by using different values for each network. 
%
Next, there are two downchirp symbols used for frequency synchronization. 
Afterwards a silence time of 0.25 symbol time is used by the receiver to align in time.
%
%Optionally, the end of preamble can include another two unmodulated base chirps that will be used for fine time and frequency synchronization. 
%
%The structure of the preamble is shown in Figure X, while 
%
%
Fig.~\ref{fig:lora_spectrum1} shows the spectrum capture of a LoRa packet with the upchirps at the beginning of preamble clearly visible. 
%
%Note, that the receiver should know the SF in advance in order to detect the preamble, as preamble size scales with SF and there is no single preamble for all SFs.
%
%Optionally, the preamble is followed by a physical header which contains payload length, FEC code rate used for payload and CRC of the header. 
%
%The header is always protected with FEC with highest code rate of 4/8. 
%
%If these three parameters are known in advance, the header can be removed completely. 
%
%This decreases the time on air of the packet. 
%
%In this case, the implicit header mechanism is applied, where the header parameters are fixed beforehand at both ends, the receiver and the transmitter side. 
%
%The payload contains either LoRaWAN MAC layer control packets or data packets. 
%
%Optionally, the payload can be followed by a payload CRC. 
%
%The frame structure is shown in Figure XYZ.

%
%Gray coding, Interleaving, Whitening, Hamming Coding

%
%
%Duty cycle restrictions
%
%No strict duty cycle restrictions
%

%The LoRaWAN medium access control (MAC) protocol is an open source protocol standardized
%by the LoRa Alliance [7] that runs on top of LoRa [1] physical layer. The LoRaWAN MAC layer
%provides the medium access control mechanism that enables communication between multiple devices
%and network gateway(s).

\begin{figure}
\centering
\includegraphics[width=1\linewidth]{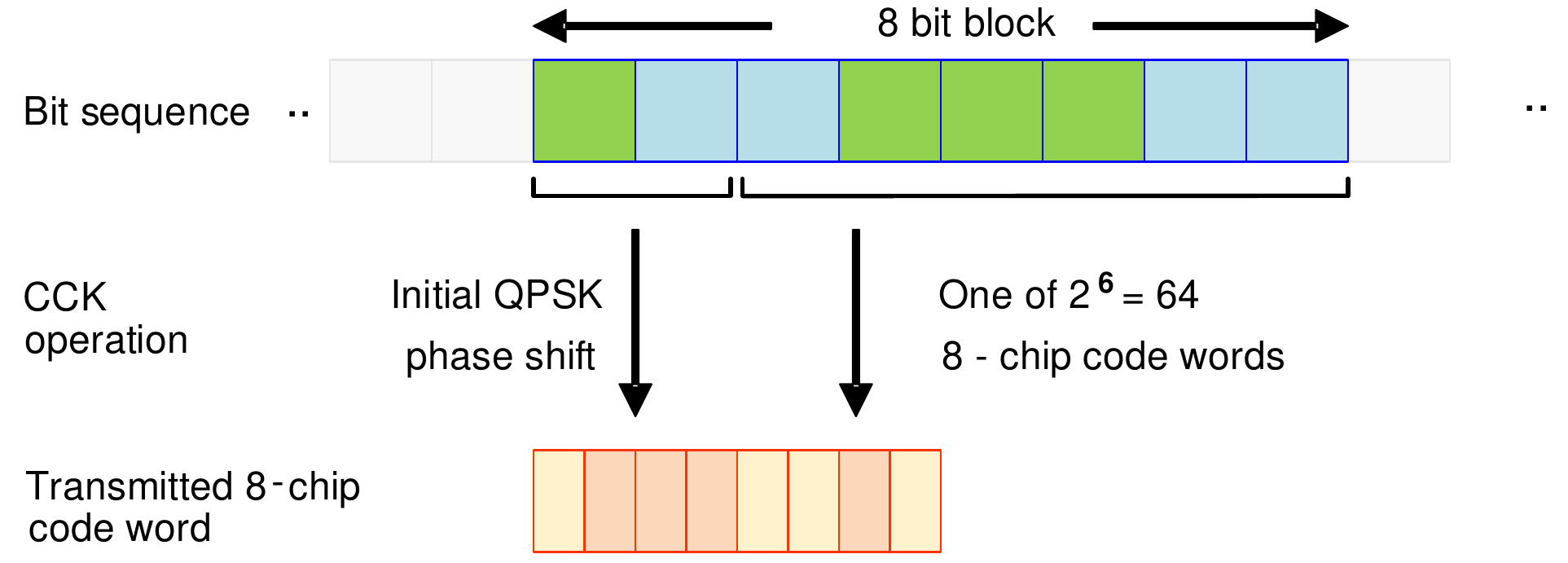}
\vspace{-5pt}
\caption{802.11b transmission at 11\,Mbit/s.}
%\vspace{-0.5em}
\label{fig:80211b_cck_11}
\end{figure}
\subsection{802.11b Primer}
%
%How 802.11b DSSS works: https://en.wikipedia.org/wiki/Complementary_code_keying
%
An IEEE 802.11b radio running at speed of 5.5 or 11\,Mbit/s is using complementary code keying (CCK) as modulation scheme~\cite{van1999new}.
CCK was adopted to supplement the Barker code used by 802.11b at 1/2\,Mbit/s to achieve data rates at the expense of slightly shorter communication distance. 
This is achieved by having the shorter chipping sequence in CCK (8\,bits versus 11\,bits in Barker code) resulting in less spreading and hence higher data rate.
%
%but more susceptible to narrowband interference resulting in shorter radio transmission range. 
%
In addition to shorter chipping sequence, CCK also uses more chipping sequences to encode more bits (4 and 8 chipping sequences for 5.5\,Mbit/s and 11\,Mbit/s respectively) and hence resulting in increased data rate.
Note, that the Barker code has only a single chipping sequence.

The CCK modulation used by 802.11b transmits data in symbols of eight chips, where each chip is a complex QPSK bit-pair at a chip rate of 11\,Mchip/s. 
In 5.5 Mbit/s and 11 Mbit/s modes respectively 4 and 8 bits are modulated onto the eight chips of the symbol $c_0,\ldots,c_7$, where~\cite{van-nee1999new}:
\begin{align*}
\bold{c} =& (c_0,\ldots,c_7) \\
=& (e^{j(\phi_1+\phi_2+\phi_3+\phi_4)},e^{j(\phi_1+\phi_3+\phi_4)},e^{j(\phi_1+\phi_2+\phi_4)},\\
& -e^{j(\phi_1+\phi_4)},e^{j(\phi_1+\phi_2+\phi_3)},e^{j(\phi_1+\phi_3)},e^{j(\phi_1+\phi_2+\phi_3+\phi_4)},\\
& -e^{j(\phi_1+\phi_2)},e^{j\phi_1})
\end{align*}
and $\phi_{1},\ldots ,\phi_{4}$ are determined by the bits being modulated.
Hence, the phase change $\phi_{1}$ is applied to every chip, $\phi_{2}$ is applied to all even code chips (starting with $c_{0}$), $\phi_{3}$ is applied to the first two of every four chips, and $\phi_{4}$ is applied to the first four of the eight chips.
Fig.~\ref{fig:80211b_cck_11} illustrates the CCK operation in 802.11b transmission at 11\,Mbit/s.
%

% -------------- Section end marker --------------
%                _       _
%               ( )_    ( )
%    ___  _   _ | ,_)   | |__     __   _ __   __
%  /'___)( ) ( )| |     |  _ `\ /'__`\( '__)/'__`\
% ( (___ | (_) || |_    | | | |(  ___/| |  (  ___/
% `\____)`\___/'`\__)   (_) (_)`\____)(_)  `\____)
%
% -------------- Section end marker --------------

\section{Wi-Lo}\label{sec:wilo}
\begin{figure}
\centering
\includegraphics[width=0.90\linewidth]{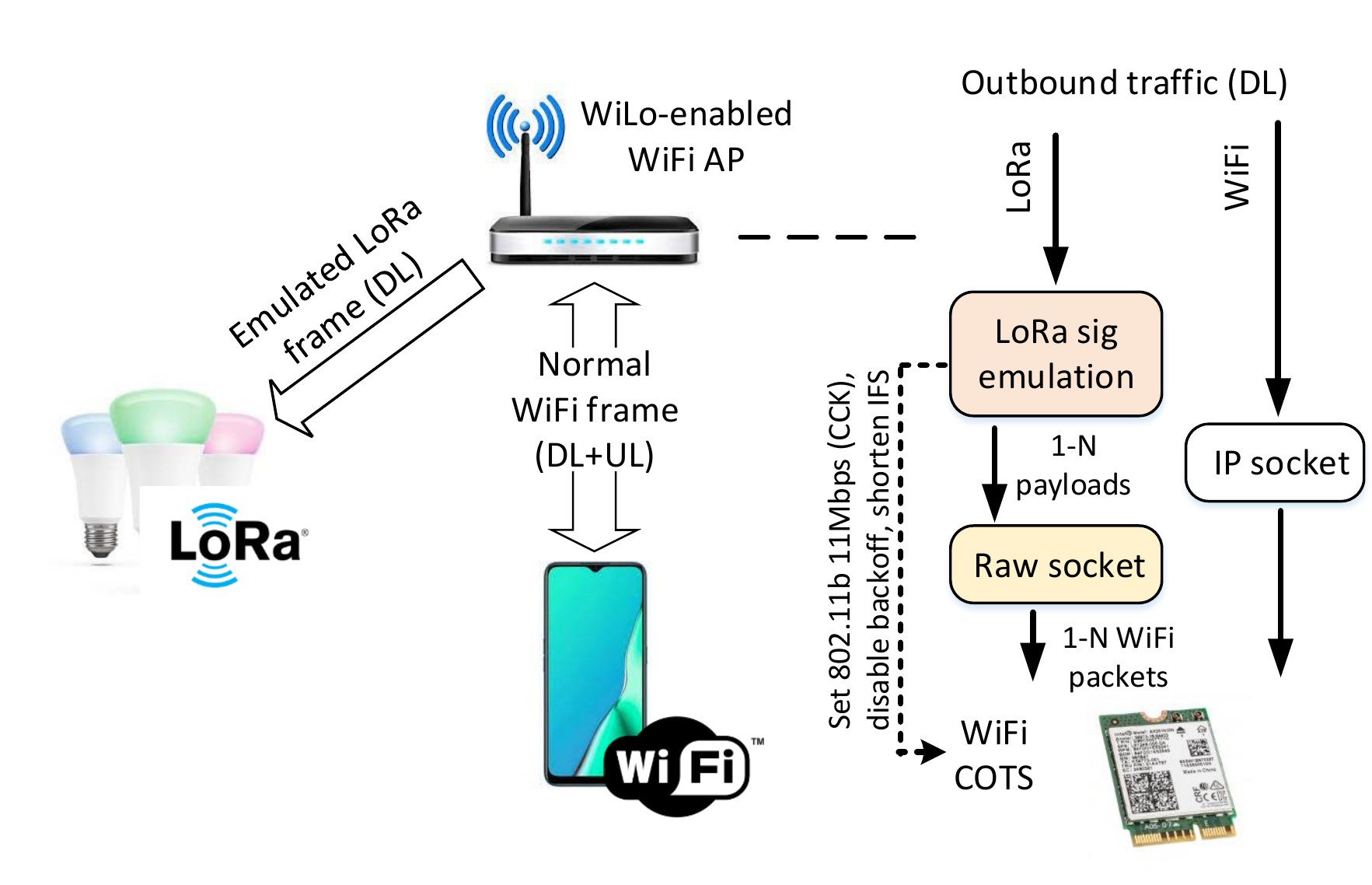}
\vspace{-5pt}
\caption{Architecture of Wi-Lo.}
%\vspace{-0.5em}
\label{fig:arch}
\end{figure}

A standard smartphone needs a multi-technology gateway (MTG) in order to be able to communicate with LoRa device in its proximity.
With Wi-Lo, such an expensive MTG that is equipped with both WiFi and LoRa radios is no longer needed.
Wi-Lo enables a commodity WiFi AP, residential or enterprise, to transmit data packets towards narrow-band IoT devices by emulating the LoRa waveform (Fig.~\ref{fig:arch}).
Therefore, no additional (radio) hardware is needed, as Wi-Lo is a pure software solution.
Instead, the Wi-Lo-enabled WiFi AP uses signal emulation to craft a downlink WiFi frame whose waveform emulates a valid LoRa packet.
Such an emulated LoRa packet can be received by unmodified commodity LoRa devices.
Note, in order to support uplink transmissions from LoRa devices to WiFi AP Wi-Lo could be extended by the approach proposed by Liu et al.~\cite{liu2020xfi}. 

\subsection{Emulating LoRa using 802.11b}
The basic idea of Wi-Lo is to use signal emulation technique where the LoRa waveform is emulated by carefully selecting the payload of a 802.11b 11 Mbps (CCK) packet transmissions so that the resulting WiFi waveform emulates a 2.4\,GHz LoRa waveform.
The basic emulation algorithm works as follows: 
\begin{enumerate}
\item Generate baseband waveform $S$ with sampling frequency $f_s = 11$\,MHz of the LoRa packet to be transmitted.
\item Process $S$ from left to right in chunks $s$ of $\Delta t=\frac{8}{11}\,\mu$s duration.
\item For each chunk $s$ find the CCK symbol $\tilde{c} \in C_{\mathrm{cck}}$ which is matching closest, i.e., has similar waveform:
\begin{align}
\tilde{c} &= \argmax_{c \in C_{\mathrm{cck}}} \Re(s) \ast \Re(c) + \Im(s) \ast \Im(c)
\end{align}
where $C_{\mathrm{cck}} = \mathrm{spread}(\mathrm{permn}([0, \frac{\pi}{2}, \pi, \frac{3\pi}{2}], 4))$, $\mathrm{permn}(\cdot)$ computes the permutations with repetition, $\mathrm{spread}(\cdot)$ is the CCK spreading as defined in 802.11b 11\,Mbps, $\ast$ is cross-correlation and $\Re()$ and $\Im()$ represent the real and imaginary part of the complex signal. Add $\tilde{c}$ to list $\tilde{C}$.
\item Derive from $\tilde{C}$ the WiFi payload bits $P$ by inverting the CCK modulation process (cf.\ Fig.~\ref{fig:80211b_cck_11}).
\item Transmit 802.11b (CCK) 11\,Mbps packet with payload given by $P$. Note, by injecting the packet using a raw socket the 802.11b preamble with correct field length is created by the WiFi driver.
%Prepend valid 802.11b preamble with length field set to be long enough to fully cover $\tilde{C}$, i.e. $\tilde{S} = [\mathrm{PR} \ \tilde{C}]$
%\item Transmit waveform $\tilde{S}$.
\end{enumerate}

\begin{figure}
\centering
\includegraphics[width=0.90\linewidth]{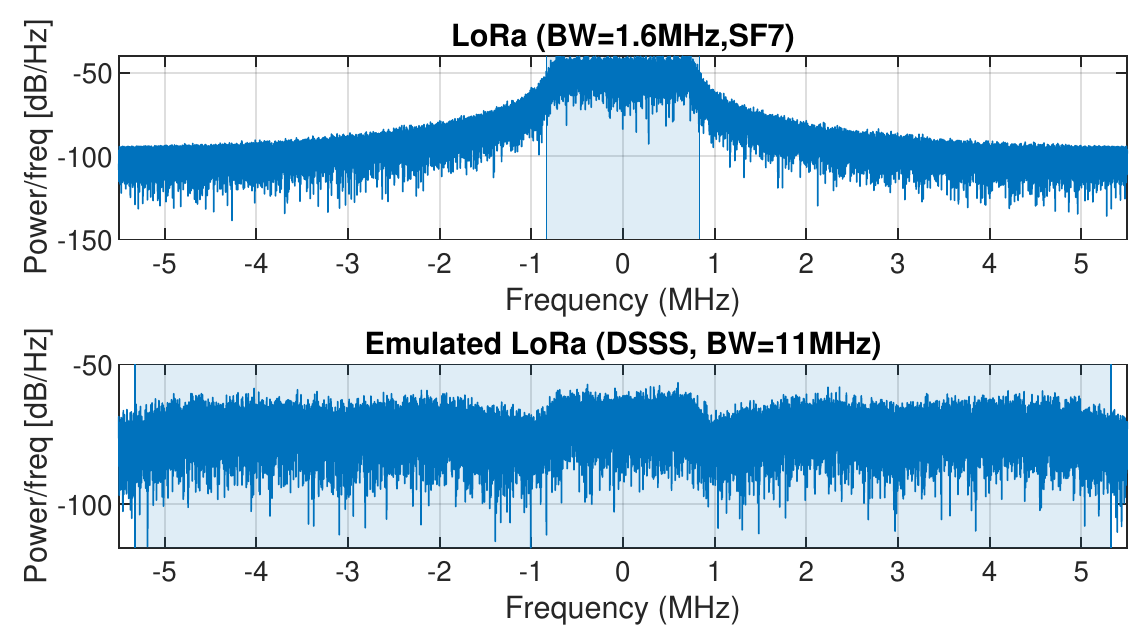}
\vspace{-5pt}
\caption{Spectrum of classical vs. emulated LoRa (Wi-Lo).}
%\vspace{-0.5em}
\label{fig:emul_lora_spectrum}
\end{figure}

Fig.~\ref{fig:emul_lora_spectrum} shows the spectrum of the normal LoRa with bandwidth of 1.6\,MHz (SF7) vs.\ the spectrum of the corresponding emulated LoRa waveform using the proposed CCK emulation.
We can clearly see the centered 1.6\,MHz LoRa signal. %\todo{but at much lower power. what's the impact?}
Moreover, the additional not needed frequency components introduced by the CCK-based emulation outside the bandwidth of 1.6\,MHz.
The resulting power loss is not an issue as the transmit power of 2.4\,GHz LoRa is limited to 8\,dBm whereas the limit for 802.11b CCK is 18\,dBm.
%
% https://wlan1nde.wordpress.com/2014/11/26/wlan-maximum-transmission-power-etsi/#:~:text=There%20exists%20two%20EIRP%20power,(63%20mW)%20for%20CCK.
%
An example for WiFi 802.11b frame emulating LoRa can be downloaded from \url{https://bit.ly/3y1f0my}\footnote{For injection you can use software like scapy. Make sure to fix the WiFi data rate to 11\,Mbps and to use the correct WiFi channel}.

\subsection{Support for Large LoRa Frames}
The transmission of multiple WiFi packets is required in order to emulate even a single LoRa packet in case the LoRa packet's airtime is larger than the maximum duration of an 802.11b transmission.
This is not unusual as even a LoRa packet with small payload can have a duration of hundreds of milliseconds when transmitted with either low bandwidth, high SF or low code rate.
Note, that the maximum data payload size of 802.11b frame is limited to 4095\,Byte, which corresponds to a transmission duration of only around 3\,ms when transmitted at 11\,Mbps.
Moreover, improvements from newer 802.11 standards like TXOP cannot be used.
%
%A major limitation of Wi-Lo so far is that the LoRa packet must be fit into a single 802.11b DSSS transmission which is problematic if you want to transmit LoRa packets with higher SF or larger payload.
%

\begin{figure}
\centering
\includegraphics[width=1.00\linewidth]{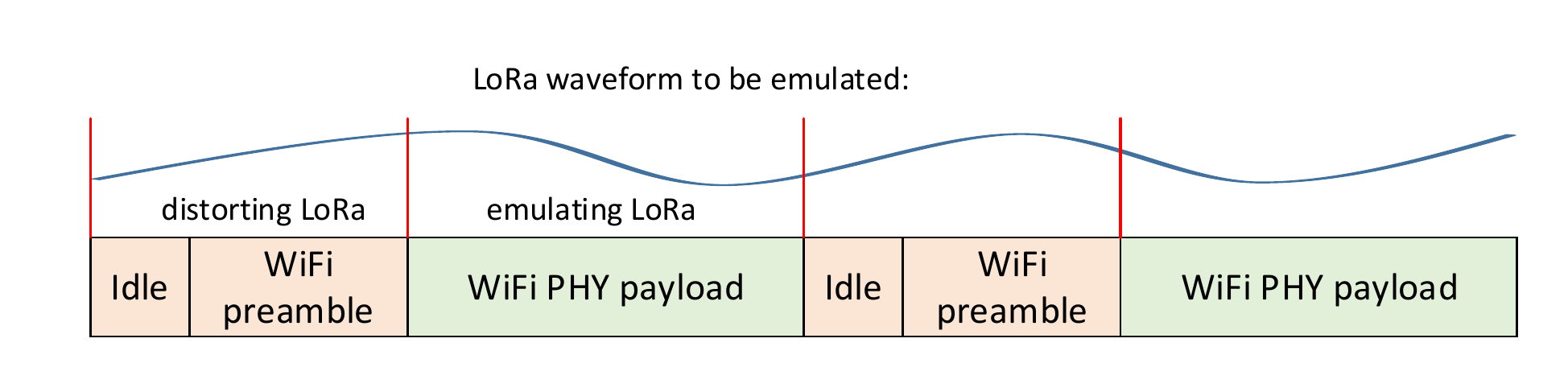}
\vspace{-5pt}
\caption{The fixed part (red) of the WiFi transmission creates distortions to the emulation process.}
%\vspace{-0.5em}
\label{fig:jumbo_frames}
\end{figure}

Wi-Lo solves this problem by emulating a single LoRa packet with a train of multiple 802.11b frames where each WiFi frame is emulating a part of the entire LoRa waveform (Fig.\ \ref{fig:jumbo_frames}).
In the worst-case, i.e. low bandwidth and high SF, already a single LoRa chirp is emulated with multiple WiFi frames.
However, such an approach is challenging as WiFi nodes have to access the channel in random fashion using carrier sensing, i.e., listen-before-talk.
Thus, WiFi transmissions are discontinuous as idle times due to interframe spaces (IFS) and backoff are inserted.
Wi-Lo solves this by disabling backoff operation and also reducing the IFS.
With Atheros AR928x NIC we were able to fully disable backoff operation while the IFS was reduced to 12\,$\mu$s $\pm$ 3\,$\mu$s. 
Another issue is that each WiFi frame starts with a preamble, 196/96\,$\mu$s in case long/short preamble, which cannot be changed or disabled.
Both the idle times between WiFi frames, i.e. IFS, as well as the WiFi preambles create distortions in the signal emulation process, i.e., $\frac{108\,\mu s}{2978\,\mu s}\approx$3.6\% of the WiFi frame duration cannot be used for the purpose of emulation (cf.\ Fig.~\ref{fig:jumbo_frames}).
Interestingly, as we will see later from our experiments, those distortions do not have a large impact on the LoRa transmission as they can be corrected by the receiver especially when long-range communication parameters, i.e., large SF and low code-rate, are used. %\todo{i.e., reducing data rates, we need to comment on that}
Moreover, a small bandwidth, e.g. 200\,kHz, makes the LoRa transmission very long in time compared to the duration of distortion thus reducing their impact.

% -------------- Section end marker --------------
%                _       _
%               ( )_    ( )
%    ___  _   _ | ,_)   | |__     __   _ __   __
%  /'___)( ) ( )| |     |  _ `\ /'__`\( '__)/'__`\
% ( (___ | (_) || |_    | | | |(  ___/| |  (  ___/
% `\____)`\___/'`\__)   (_) (_)`\____)(_)  `\____)
%
% -------------- Section end marker --------------

\subsection{Prototype Implementation}\label{sec:proto}
The Wi-Lo prototype was implemented using COTS hardware for both WiFi and LoRa.
On the WiFi side we used Atheros AR928x (802.11n) NIC, which still support the old 802.11b standard.
For 2.4\,GHz LoRa we used the radio module iM282A from IMST Wireless Solutions.
It is a bidirectional radio module based on the transceiver SX1280 of Semtech. % supporting LoRa modulation technology in 2.4\,GHz frequency band.
The chip's sensitivity is up to -130\,dBm (SF12, BW=200\,kHz) while the maximum transmit power is 8\,dBm resulting in a link budget of more than 138\,dB.
%
%Fig. X shows the setup.

% -------------- Section end marker --------------
%                _       _
%               ( )_    ( )
%    ___  _   _ | ,_)   | |__     __   _ __   __
%  /'___)( ) ( )| |     |  _ `\ /'__`\( '__)/'__`\
% ( (___ | (_) || |_    | | | |(  ___/| |  (  ___/
% `\____)`\___/'`\__)   (_) (_)`\____)(_)  `\____)
%
% -------------- Section end marker --------------

\section{Performance Evaluation}\label{sec:results}
We evaluated the performance of the Wi-Lo prototype in a mixed environment, i.e., office space and outdoor. %\todo{that is sender in office, receiver somewhere outdoors? what about indoor-indoor and outdoor-outdoor?}
%
%We configured Wi-Lo to emulate a LoRa frame with BW=1.6\,MHz, SF6, code rate 4/5 and a payload of one Byte.
%
%The LoRa frame's duration is 2\,ms which is short enough so that it fits into a single DSSS WiFi frame.
%
The 2.4\,GHz ISM band on channel 6 (2427\,MHz) was used.
During the experiment, a COTS WiFi node with Atheros WiFi NIC was sending valid 802.11b CCK (11\,Mbps) frames.
The WiFi payload of single/multiple WiFi frames was precomputed in Matlab to emulate the LoRa frame as described in Section~\ref{sec:wilo}.
%
%The resulting WiFi frame had a frame size of X Bytes and a duration of X\,ms.
%
On the LoRa side, we used the COTS LoRa chip in packet sniffer mode in order to collect all packets.
For each received packet the signal strength (RSSI in dBm) and the SNR was reported and collected.
As baseline for comparison a pure LoRa setup was used, i.e. also the transmitter was a COTS LoRa chip.

\subsection{Initial Over-the-Cable Tests}
\begin{figure}
\centering
\includegraphics[width=0.80\linewidth]{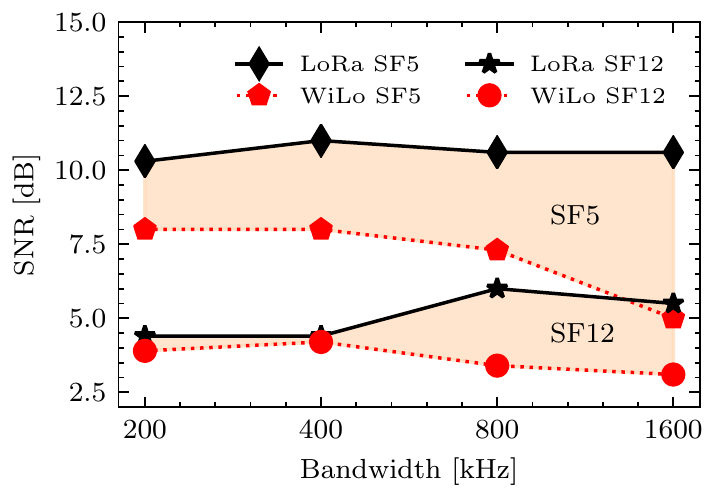}
\vspace{-5pt}
\caption{Transmission over-cable.}
%\vspace{-0.5em}
\label{fig:bw_vs_snr_cable}
\end{figure}

In order to understand the performance of Wi-Lo's signal emulation capabilities we performed experiments over coax cable with 30\,dB attenuator. % with different attenuations so that receive signal strength (RSSI) was between -70\,dBm and -40\,dBm.
Such a configuration represents the operation at high signal strength, i.e., RSSI at around -25\,dBm, mimicking perfect channel conditions.
We compared Wi-Lo with a pure LoRa setup with respect to the reported SNR value at the LoRa receiver side.
The results are shown in Fig.~\ref{fig:bw_vs_snr_cable}.
Although the RSSI was the same for both configurations, the SNR was different.
For SF5, we see a SNR drop between 2.3 and 5.5\,dB depending on the bandwidth used by LoRa.
For larger bandwidth, the SNR drop is larger.
This can be explained by the fact that emulating a LoRa signal with larger bandwidth is harder than one with lower, i.e., the probability is smaller to find a CCK chip sequence which perfectly fits a given part of the LoRa waveform.
With SF12, the situation is similar but the SNR drop is smaller, i.e., max 2.5\,dB.
We can conclude that the SNR drop due to imperfections of the CCK-based emulation process is smaller for LoRa transmissions using small bandwidth, preferably 200\,kHz, and larger spreading factor, preferably SF12.
From the practical point of view the SNR drop plays a minor role as the required SNR for LoRa is very small, i.e. -2.5\,dB and -20\,dB for SF5 and SF12 respectively.
%
%\todo{well, SF12 is more robust but slower, right, SF5 is much more impacted, particularly at 1600kHz, we should at least already now indicate whether this is relevant and provide a forward reference to experiments showing the impact}

%, i.e. 12\,dB and 4\,dB for pure LoRa and Wi-Lo respectively.
%
%The SNR drop of 8\,dB is due to the approximation of LoRa's 1.6\,MHz signal using the proposed DSSS based signal emulation (cf. Fig.\ref{fig:emul_lora_spectrum}).
%
%Fig.~\ref{fig:bw_vs_snr_cable}: SNR drop of 2.5\,dB and 5.5\,dB for SF 12 and SF=5 respectively.

\subsection{Emulation of Large LoRa Frames}
In Section~\ref{sec:wilo}, we mentioned that Wi-Lo is able to emulate LoRa frames even if their airtime is longer than the one of a single 802.11b frame.
This is achieved by sending multiple WiFi frames each emulating a part of the entire LoRa waveform.
Unfortunately, the idle times between WiFi frames as well as the WiFi preambles create distortions in the signal emulation process.
Therefore, we examined this impact in more detail.
Wi-Lo was configured to emulate a LoRa frame using SF12 with code rate 4/8 and a bandwidth of 200\,kHz.
The airtime of such a LoRa frame was 813\,ms.
Such a long frame was emulated using 283 WiFi frames each 2.8\,ms long.
%
%The distortion introduced is due to idle periods between wifi frames of 13\,mus\footnote{This is the gap between WiFi frames we achieved with COTS WiFi Atheros HW with disabled backoff and CCA.} and wifi short preamble of 98\,mus.
%
The Wi-Lo transmitter was connected over coax cable to the LoRa receiver.
%Setup: WiFi frames send out of USRP connected over coax-cable to LoRa receiver.
%
Comparison with the baseline show a SNR degradation of only $\approx 0.7$\,dB. %\todo{any results figure?}
Therefore, we can conclude that emulation of long LoRa frames is feasible.

\subsection{Over-the-air Tests}
Next, we performed experiments with over-the-air transmissions.
The Wi-Lo transmitter was fixed and placed indoors while the LoRa receiver was mobile.
While walking around with the receiver indoors and outdoors the LoRa packets were received.
The walking area was mostly outdoors and within a radius of 250\,m around the transmitter.
For each received packet the RSSI and the SNR as reported by LoRa chip was collected.
The following configuration was used for LoRa: BW=1.6\,MHz, SF6, code rate 4/5 and a payload of one Byte.
The LoRa frame's duration was 2\,ms which is short enough so that it fits into a single 802.11b frame.

\begin{figure}
\centering
\includegraphics[width=0.85\linewidth]{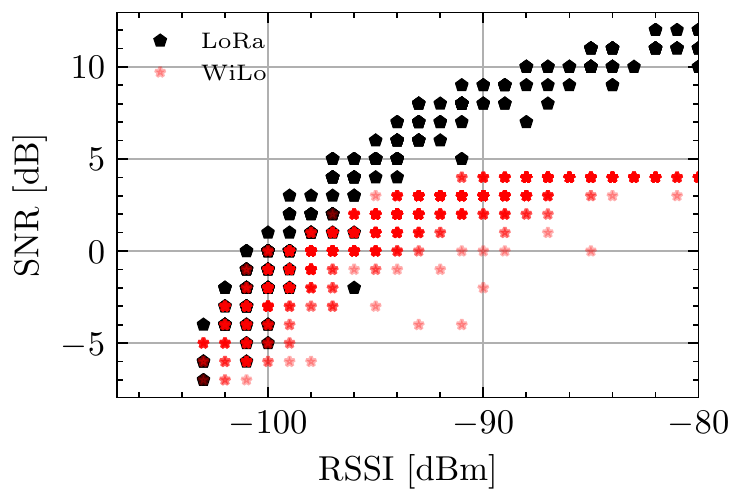}
\vspace{-5pt}
\caption{Transmission over-the-air (SF6).}
%\vspace{-0.5em}
\label{fig:lora-air}
\end{figure}

We compared Wi-Lo with the baseline using pure LoRa devices (Fig.~\ref{fig:lora-air}). %\todo{what's the color coding here (light red to dark red)?}
As in the previous wired experiment, we see SNR degradation due to emulation. %, i.e. with Wi-Lo the SNR was 8\,dB lower as 
%the existence of an SNR wall which is around 4-5\,dB for Wi-Lo which cannot be passed even at very high RSSI. % the SNR does not go above 4-5\,dB.
%
%This is due to signal impairments on the transmitter side due to emulation.
%
However, the reduction in SNR is not a problem in real setups as the long-range signal reception is mostly limited by noise, i.e., weak signal.
This is confirmed by our results where we see no difference in the SNR of both pure LoRa and Wi-Lo when operating at very weak signal levels.
The lowest RSSI for which LoRa packets were correctly received was -103\,dBm, which was the same for both baseline and Wi-Lo.
%
%This is comparable to baseline where standard LoRa transmitter is used.
%
%SNR loss not an issue if operating at low signal strength as performance is limited by noise.

\subsection{Distance Measurements}
\begin{figure}
\centering
\includegraphics[width=0.90\linewidth]{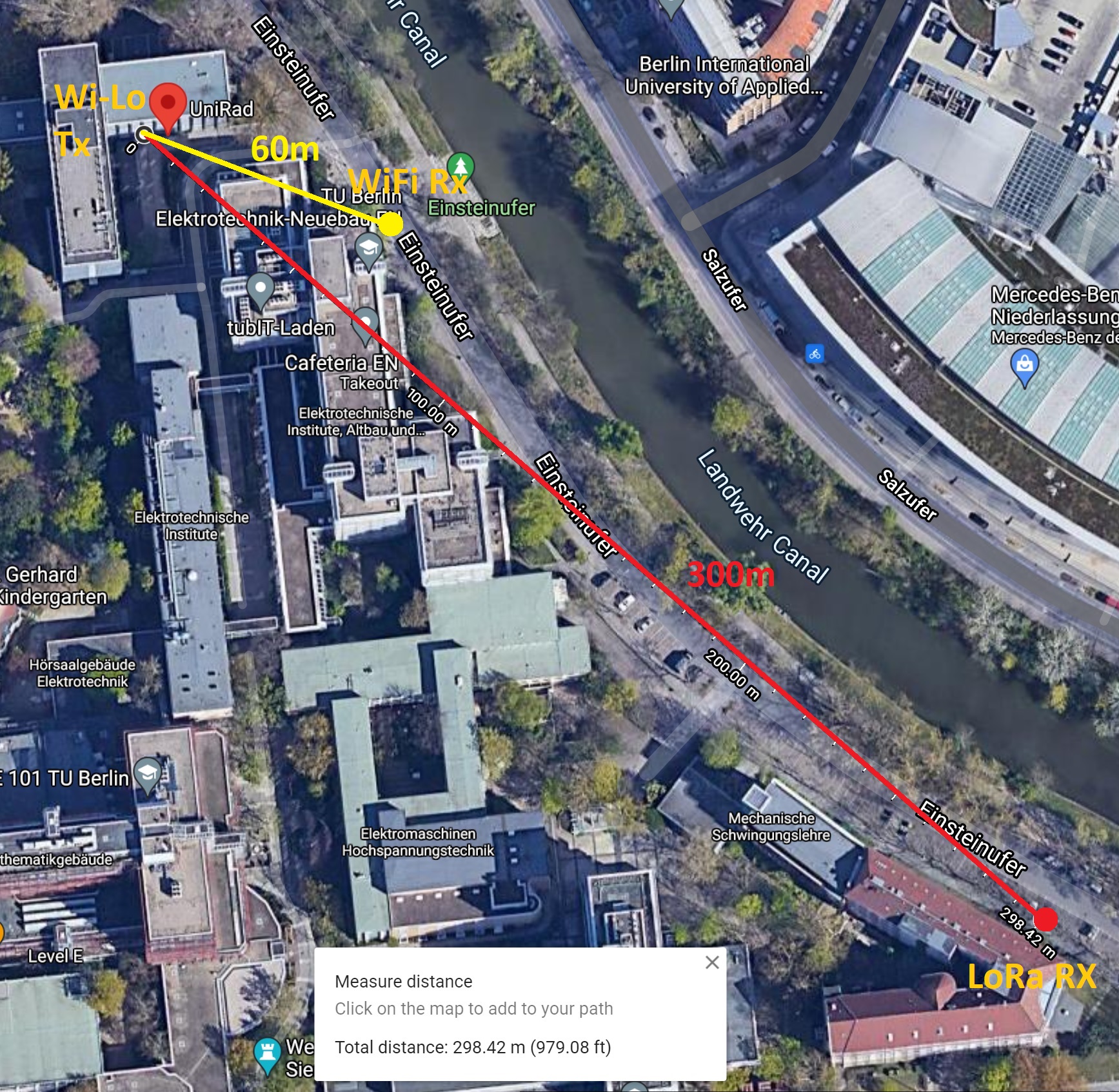}
%\vspace{-1pt}
\caption{Long-distance measurements.}
%\vspace{-0.5em}
\label{fig:wilo-dist}
\end{figure}

Finally, we performed outdoor experiments on the campus of TU Berlin to find out the maximum communication distance, which can be achieved with Wi-Lo.
Wi-Lo transmitter was placed at the window board of our building (10\,m above the ground) and configured to send emulated LoRa frames with bandwidth of 1.6\,MHz, SF6, code rate 4/5 (Fig.~\ref{fig:wilo-dist}).
As we wanted to analyze the difference in communication range between LoRa and WiFi we used two different receivers: i) a COTS LoRa receiver and ii) a COTS WiFi receiver.
During the measurements, we walked away from the transmitter on the sidewalk along the river.
The propagation characteristic was NLOS all the time, i.e., the large building was blocking the LOS path.

Fig.~\ref{fig:wilo-dist} shows the maximum distance at which the receiver was able to receive frames from the transmitter.
In case of the normal COTS WiFi the maximum distance was 60\,m to receive the 11\,Mpbs encoded frames.
The distance of the LoRa receiver was $5\times$ larger, i.e., 300\,m.
This shows again that the SNR drop due to emulation has no visible impact on the maximum communication range.
Note, higher distances can we achieved when using larger SF values. %\todo{this is all with a Wifi sender and a LoRa receiver, right? shouldnt this be compared between LoRa sender vs WiFi sender and a LoRa receiver?}

% -------------- Section end marker --------------
%                _       _
%               ( )_    ( )
%    ___  _   _ | ,_)   | |__     __   _ __   __
%  /'___)( ) ( )| |     |  _ `\ /'__`\( '__)/'__`\
% ( (___ | (_) || |_    | | | |(  ___/| |  (  ___/
% `\____)`\___/'`\__)   (_) (_)`\____)(_)  `\____)
%
% -------------- Section end marker --------------

%\section{Discussion}\label{sec:discussion}
%
%The proposed approach has several limitations.
%
%A LoRa transmission with bandwidth of 200-1600\,kHz is converted into a WiFi transmission occupying 22\,MHz of spectrum which is spectrally inefficient.
%
%However, such bandwidth expansion is in scenarios with high contention and interference from WiFi helpful as each emulated LoRa packet is preceding with a valid 802.11 preamble hence triggering clear-channel assessment (CCA) operation in the co-located WiFi nodes.
%
%Moreover, we can use the virtual channel reservation used by WiFi to make LoRa transmissions more robust against WiFi interference.
%
%This could be easily achieved by exploiting WiFi's virtual channel reservation capabilities (e.g., CTS-to-self frame).

%\todo{not sure if this should stay in an early arXiv version. this is all very relevant for teh complete story - and paper}

% -------------- Section end marker --------------
%                _       _
%               ( )_    ( )
%    ___  _   _ | ,_)   | |__     __   _ __   __
%  /'___)( ) ( )| |     |  _ `\ /'__`\( '__)/'__`\
% ( (___ | (_) || |_    | | | |(  ___/| |  (  ___/
% `\____)`\___/'`\__)   (_) (_)`\____)(_)  `\____)
%
% -------------- Section end marker --------------

\section{Related Work}\label{sec:relatedwork}
An overview on cross-technology communication (CTC) for IoT was given by Chen et al.~\cite{chen2019survey}.
The power of OFDM/QAM-based signal modulation was shown in past for the case of CTC between WiFi and ZigBee~\cite{li2017webee,chen2018twinbee,li2018longbee}, WiFi and Bluetooth~\cite{jiang2017bluebee}, WiFi and LTE~\cite{gawlowicz2018enabling, gawlowicz2020punched}.
With Wi-Lo, we show that CCK-based signal modulation is also feasible using the example of 2.4\,GHz LoRa.
Hence, a commodity WiFi device can send LoRa packets towards IoT devices supporting the LoRa protocol.
%\todo{more details needed}
%
Liu~\cite{liu2020xfi} has shown that COTS WiFi hardware can be used for the reception of LoRa transmissions.
This is achieved using a technique called signal hitchhiking, i.e., when a smartphone is receiving a WiFi packet from an AP, IoT devices transmit simultaneously, leading to intentional collisions with the WiFi packet in the air.
In this way, the LoRa data hitchhikes on the WiFi packet and enters the WiFi radio where it can be decoded through waveform reconstruction and subsequent LoRa decoding.
In this paper, we show for the first time that with signal emulation a WiFi sender is able to send LoRa frames.
%\todo{this paper: first to show that a WiFi sender can be used to send LoRa packets}

% -------------- Section end marker --------------
%                _       _
%               ( )_    ( )
%    ___  _   _ | ,_)   | |__     __   _ __   __
%  /'___)( ) ( )| |     |  _ `\ /'__`\( '__)/'__`\
% ( (___ | (_) || |_    | | | |(  ___/| |  (  ___/
% `\____)`\___/'`\__)   (_) (_)`\____)(_)  `\____)
%
% -------------- Section end marker --------------

\section{Conclusion}\label{sec:conclusion}
We presented Wi-Lo that allows to convert a commodity WiFi AP into an IoT gateway supporting 2.4\,GHz LoRa protocol with a simple software update.
Wi-Lo uses signal emulation technique based on 802.11b CCK modulation in order to emulate a downlink LoRa transmission.
This allows a WiFi-enabled smartphone to use the Wi-Lo gateway in order to send packets to LoRa compliant IoT devices.
As future work, we plan to support other IoT protocols like Bluetooth LE.

% -------------- Section end marker --------------
%                _       _
%               ( )_    ( )
%    ___  _   _ | ,_)   | |__     __   _ __   __
%  /'___)( ) ( )| |     |  _ `\ /'__`\( '__)/'__`\
% ( (___ | (_) || |_    | | | |(  ___/| |  (  ___/
% `\____)`\___/'`\__)   (_) (_)`\____)(_)  `\____)
%
% -------------- Section end marker --------------

\bibliographystyle{IEEEtran}
\bibliography{references,IEEEabrv}

\end{document}